\documentclass[page-classic]{epl2}
\usepackage{hyperref}

\title{Interaction between \emph{Mn}  Ions and Free Carriers in Quantum Wells with Asymmetrical Semimagnetic Barriers}
\shorttitle{Magnetic interaction in QWs} 

\author{D.~M.~Zayachuk\inst{1} \and T.~Slobodskyy\inst{2,3}, G.~V.~Astakhov\inst{3,4}, C.~Gould\inst{3}, G.~Schmidt\inst{3,5}, W.~Ossau\inst{3}, L.~W.~Molenkamp\inst{3}}

\institute{
  \inst{1} Lviv Polytechnic National University, 12 Bandera St, 79013 Lviv, Ukraine\\
  \inst{2} Present address: Institute for Synchrotron Radiation, Karlsruhe Institute of Technology, 76344 Eggenstein-Leopoldshafen, Germany\\
  \inst{3} Physikalisches Institut der Universit\"{a}t W\"{u}rzburg, 97074 W\"{u}rzburg, Germany\\
  \inst{4} A. F. Physico-Technical Institute, Russian Academy of Sciences, 194021 St. Petersburg, Russia\\
  \inst{5} Present and permanent address: Institut f\"{u}r Physik, Universit\"{a}t Halle, 06099 Halle, Germany
}

\pacs{78.55.Et}{Photoluminescence, properties and materials, II-VI semiconductors}
\pacs{73.21.Fg}{Electron states and collective excitations in quantum wells}
\pacs{78.20.Ls}{Optical properties of bulk materials and thin films, Magneto-optical effects}
\pacs{71.55.Gs}{Impurity and defect levels, II-VI semiconductors }
\pacs{73.61.Ga}{Electrical properties of specific thin films, II-VI semiconductors}

\abstract{Investigations of photoluminescence (PL) in the magnetic field of quantum structures based on the $ZnSe$ quantum well with asymmetrical $ZnBeMnSe$ and $ZnBeSe$ barriers reveal that the introduction of $Be$ into semimagnetic $ZnMnSe$ causes a decrease of the exchange integrals for conductive and valence bands as well as the forming of a complex based on $Mn$, degeneration of an energy level of which with the energy levels of the $V$ band of $ZnBeMnSe$ or $ZnSe$ results in spin-flip electron transitions.}

\begin{document}

\maketitle

In the last ten years, semimagnetic quantum structures attracted interest for both fundamental and practical applications, especially for spin electronics \cite{SSQC, SPS, Fie99, 1}. Both structures with magnetic ions (usually $Mn^{2+}$) in potential wells and barriers have been studied \cite{Cro95, 2, 3}. In the first case, magnetic interaction between the free carriers and the localized 3\emph{d} electrons of $Mn^{2+}$ ions (the so-called $s,~p-d$ exchange interaction) is stronger. It leads to an appreciable intensification of the effects caused by the magnetic field. However, in contrast to nonmagnetic structures characterized  by the long spin lifetime of carriers \cite{Kik99, Hof06, Ast08}, in semimagnetic materials the exchange interaction with the presence of magnetic impurities in the quantum well stimulates spin relaxation processes \cite{4, Ast08a}. Hence, it is preferable to separate the carriers from the magnetic media. In this case, the exchange interaction between the 2\emph{D} free carriers of the quantum well and the ions of magnetic impurities in the barrier is weaker and the exchange integral is driven by the penetration of the carrier's wave function tails into the barrier. At the same time, additional possibilities for spin manipulation appear by influencing the magnetic interaction transfer from the ferromagnetic material in the barrier to the carriers in the quantum well by technological methods. In particular, an additional non-magnetic layer introduced between the semimagnetic barrier and a non-magnetic quantum well may be used for tuning spin interaction. Here, the method will be presented using the quantum structures based on the $ZnSe$ quantum well with asymmetrical $ZnBeMnSe$ and $ZnBeSe$ barriers as an example.

\begin{figure}\
 \onefigure[width=6.5cm]{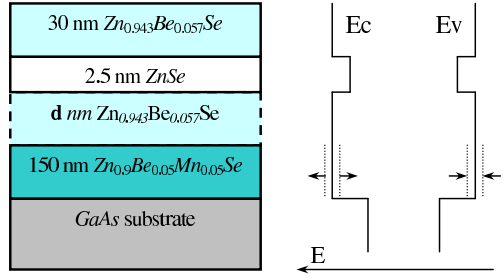}
 \caption{Schematic view of the sample layout (left) with the corresponding energy band profile (right). The spin split sub-bands of the conduction band ($E_{C}$) and the valence band ($E_{V}$) in the magnetic field are depicted by arrows. The $Zn_{0.943}Be_{0.057}Se$ barrier thickness \textbf{d} was varied as 0 (named $a1$), 2.5 ($b1$), 7.5 ($c1$), and 12.5 nm ($d1$).}
 \label{f1}
\end{figure}

The quantum structures were grown by molecular beam epitaxy on a $GaAs$ substrate. The sample layout is depicted in Fig.~\ref{f1}. The nonmagnetic space layer \textbf{d} is used for varying magnetic interaction in the structure. PL spectra at 1.2~K in magnetic fields up to 5.25~T in Faraday geometry were measured to study peculiarities of an exchange interaction between the $Mn$ ions and free carriers. For optical excitation, we used a stilbene-3 dye laser pumped by the ultraviolet lines of an Ar-ion laser. For non-resonant excitation, the laser energy is tuned to $E_{exc}$~=~2.94~eV exceeding the band gap of the $Zn_{0.9}Be_{0.05}Mn_{0.05}Se$ barrier.

\begin{figure}\
 \onefigure[width=8cm]{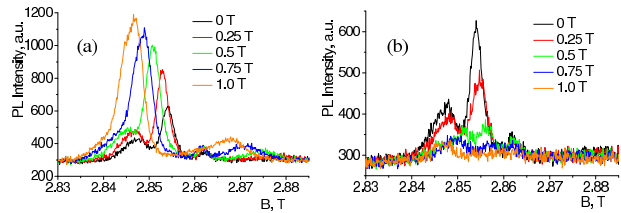}
 \caption{PL spectra of $\sigma^{+}$ (a) and $\sigma^{-}$ - polarization (b) of the 150~nm~$Zn_{0.9}Be_{0.05}Mn_{0.05}Se$/ 2.5~nm~$ZnSe$/ 30~nm~$Zn_{0.943}Be_{0.057}Se$ structure at different magnetic fields.}
 \label{f2}
\end{figure}

\begin{figure}\
 \onefigure[width=7cm]{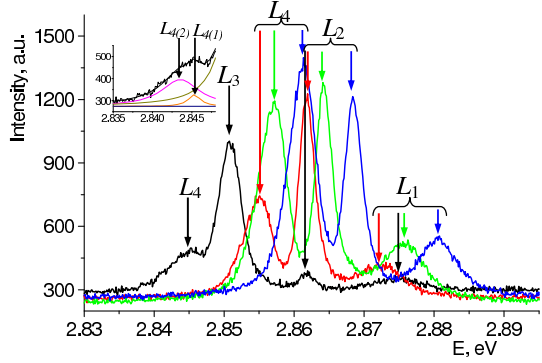}
 \caption{PL spectra of $\sigma^{+}$ polarization of the $a1$ (black line), $b1$ (red line), $c1$ (green line), and $d1$ (blue line) structures in 0.5~T magnetic fields. Inset: Reproduction of the L$_4$ band for the $a1$ structure by two Lorentz components L$_{4(1)}$ and L$_{4(2)}$.}
 \label{f3}
\end{figure}

\begin{figure}\
 \onefigure[width=6cm]{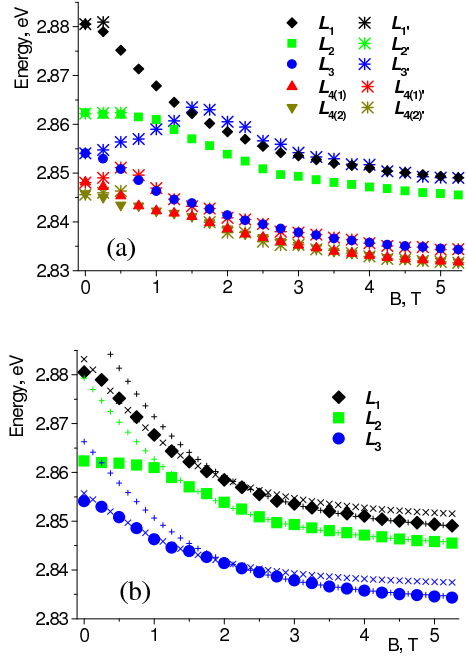}
 \caption{Field dependencies of the energy positions of the Lorentz components of PL spectra of the $a1$ structure: (a) Experimental data; (b) Comparison of experimental and calculated data for the $\sigma^{+}$ polarization bands using reconstructed energy diagram of the structure. The labels of the panel (b) show the relation of the weak (x) and strong (+) magnetic field components to the resulting magnetic field dependence curve. Contribution of the components is characterized by the ratio of 1/3. }
 \label{f4}
\end{figure}

The dynamics of the PL spectra of both $\sigma+$ and $\sigma-$ polarization of the structure without any additional nonmagnetic barrier $Zn_{0.943}Be_{0.057}Se$ between the semimagnetic layer $Zn_{0.9}Be_{0.05}Mn_{0.05}Se$ and the quantum well $ZnSe$ (the structure $a1$, Fig.~\ref{f1}) in magnetic fields up to 1~T are shown in Fig.~\ref{f2}. The effect of the additional nonmagnetic barrier $Zn_{0.943}Be_{0.057}Se$ on the PL spectra of the structures is evident in Fig.~\ref{f3} by the example of the $\sigma^{+}$ polarizations spectra in the 0.5 T magnetic fields. Obviously, it appreciably changes the PL spectra of the structure. Detailed discussion of the changes comes out of the scope of this article and will be reported elsewhere. We accent on the reduction of the luminescence bands number from four to three under the effect of the barrier since it holds the key for understanding the emission bands origin. 
The luminescence bands of the experimental PL spectra are fitted by Lorentz shape. The short-wave PL bands L$_1$ - L$_3$ are reproduced by a single Lorentz curve. Hence, we treat them as singles in further analysis. The long-wave band  L$_4$, on the other hand, requires two nearly situated Lorentz curves, and therefore is considered to be a superposition of two close located  L$_{4(1)}$ and L$_{4(2)}$ bands  (see inset in Fig.~\ref{f3}).
The dependence of the PL spectra component energy positions on the magnetic field for the $a1$ structure is shown in Fig.~\ref{f4}~(a).

As mentioned above, four bands form the PL spectra of the $a1$ structure in the absence of an external magnetic field. Their behavior in the magnetic field is absolutely different for different polarizations of radiation. The $\sigma^{+}$ polarization luminescence bands are detected throughout the entire range of magnetic fields and move to the long-wave range, if $B$ increases. It is evident from Fig.~\ref{f3} one can see that the $L_3$ luminescence band is not observed in all structures with an additional nonmagnetic barrier $Zn_{0.943}Be_{0.057}Se$ between $Zn_{0.9}Be_{0.05}Mn_{0.05}Se$ and the $ZnSe$ quantum well. For $\sigma^{-}$ polarization, the $L_{1'}$ and $L_{2'}$ high-energy bands practically disappear when $B$~$>$~0.5~T. The $L_{3'}$ and $L_{4'}$ bands can be detected adequately for any investigated magnetic field.

The source of changes in the PL spectra of the semimagnetic semiconductors under consideration in the magnetic field is the giant Zeeman splitting which is caused by $s,~p~-~d$ interaction between the localized $d$ electrons of $Mn^{2+}$ ions and the band electrons of the host material\cite{5}. Zeeman splitting of $C$ and $V$ bands in magnetic field $B$ may be written as \cite{5, 6, 7}:

\begin{equation}
\label{e1}
E(B)=E(0)\mp(\chi N_{0})\tilde{x}<S_{Z}>)
\end{equation}

where

\begin{equation}
\label{e2}
\chi N_{0}=\alpha N_{0}-\beta N_{0}
\end{equation}

and $\tilde{x}$  is the effective $Mn$ concentration, $\alpha N_{0}$ and $\beta N_{0}$ are the exchange integrals for conductive and valence bands, respectively, and $S_{Z}$ is the thermal average of the $Mn$ spin, given by:

\begin{equation}
\label{e3}
S_{Z}=\frac{5}{2}B_{5/2}[5\mu_{B}B/k_{B}(T+T_{eff})]
\end{equation}

where $B_{5/2}$ is the Brillouin function of the argument in square brackets, $\mu_{B}$ is the Bohr magneton, $k_{B}$ is the Boltzmann constant, $T$ is the temperature, and $T_{eff}$ is an empirical parameter representing antiferromagnetic interaction between the $Mn$ ions. For the calculations, parameter $T_{eff}$ is taken to be equal to 1.75~K in accordance with the empirical ratio obtained in \cite{6} for $Zn_{1-x}Mn_{x}Se$.

Eq.~\ref{e1} describes the PL band behavior in the magnetic field fairly well. The values of the $E(0)$ and $(\chi N_{0})$ parameters are different for different bands. The parameters are also different for weak and strong magnetic fields. We shall denote these parameters as $E_{iw}(0)$, $(\chi N_{0})_{iw}$ and $E_{is}(0)$, $(\chi N_{0})_{is}$, where index (i) marks the respective band, while (w) and (s) indexes mark the weak and strong magnetic field, respectively.

\begin{figure}\
 \onefigure[width=6cm]{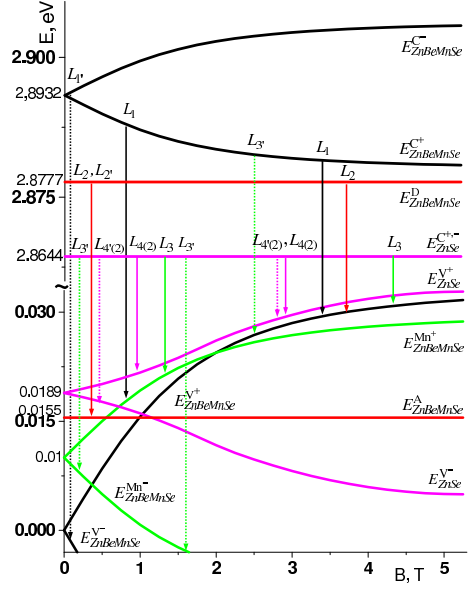}
 \caption{The positions of different energy levels of the $a1$ structure versus magnetic field. The arrows point the emitting transitions of $\sigma^{+}$ (solid arrows) and $\sigma^{-}$ polarization (dotted arrows) of the different Lorentz components of PL spectra. Only the heavy holes level which defines the valence band edge is shown for the ZnSe quantum well.}
 \label{f5}
\end{figure}

Data in Fig.~\ref{f4}~(a) coupled with values of the $E_{iw}(0)$, $(\chi N_{0})_{iw}$, $E_{is}(0)$, and $(\chi N_{0})_{is}$ parameters enable us to construct the energy diagram of the levels which form the PL spectra of the structure under consideration and their magnetic field dependence (Fig.~\ref{f5}). Fig.~\ref{f4}~(b) shows the reproduction of the experimental spectra with calculated data obtained on the basis of the energy diagram shown in Fig.~\ref{f5}.

The $L_{1}$ band in strong ($B$~$>$~2~T) magnetic fields, which has the highest emission energy, is attributed to transitions between $C$ and $V$ bands of the $Zn_{0.9}Be_{0.05}Mn_{0.05}Se$ barrier (the black lines in Fig.~\ref{f5}). $(\chi N_{0})_{1s}$ for it is equal to 0.368~eV. It is approximately three times less than the value $\alpha N_{0}-\beta N_{0}$=1.17~eV for the $Zn_{1-x}Mn_{x}Se$ solid solutions \cite{2, 7}. This implies that embedding of $Be$ in $ZnMnSe$ decreases the value of the exchange integrals of $C$ and $V$ bands in the $ZnBeMnSe$ solid solution.

The next $L_{2}$ band which practically does not depend on \emph{B} in weak magnetic fields B~$\leq$~0.5~T may be attributed to emitting transitions between the donor and acceptor levels in the $Zn_{0.9}Be_{0.05}Mn_{0.05}Se$ layer. The red lines in Fig.~\ref{f5} show these levels. The energy position of the $L_{2}$ band begins to depend on \emph{B} appreciably when the acceptor level intercepts with the $Zn_{0.9}Be_{0.05}Mn_{0.05}Se$ valence band edge, the position of which depends on the magnetic field (Fig.~\ref{f5}).

The $L_{3}$ luminescence band is identified to occur due to the following reasons: (i) The difference between the parameter values ($(\chi N_{0})_{3w}$)=0.152~eV and ($(\chi N_{0})_{1w}$)=0.264~eV coincides well with odds between the values of the parameters ($(\chi N_{0})_{1s}$)= 0.368~eV and ($(\chi N_{0})_{1w}$). (ii) The $L_{3}$ band can be extinguished by embedding an additional $Zn_{0.943}Be_{0.057}Se$ barrier between the $ZnSe$ and $Zn_{0.9}Be_{0.05}Mn_{0.05}Se$ layers. From this, we conclude that in weak magnetic fields the $L_{3}$ band is formed by an indirect transition between the 2$D$ conduction band of the $ZnSe$ quantum well and the acceptor level in the semimagnetic $Zn_{0.9}Be_{0.05}Mn_{0.05}Se$ barrier with a field-dependent energy position. The transitions on this acceptor level from the $Zn_{0.9}Be_{0.05}Mn_{0.05}Se$ conduction band form the $L_{1}$ band in weak magnetic fields. The green lines in Fig.~\ref{f5} illustrate the splitting of the probable acceptor level in the magnetic field. This level changes its position with the $Zn_{0.9}Be_{0.05}Mn_{0.05}Se$ $V$ band until approx. 2~T. Therefore, the transitions between the 2$D$ conduction band of $ZnSe$ and the $Zn_{0.9}Be_{0.05}Mn_{0.05}Se$ $V$ band form the $L_{3}$ band in the magnetic field over 2~T only.

Since the acceptor level position is sensitive to the magnetic field, we assume that the complexes incorporating $Mn$ form the acceptor centers mentioned. The high probability of existence of some complexes in $ZnBeMnSe$ follows from the relationship between the ion radiuses of $Zn$, $Mn$, and $Be$. Their difference is so large that it is impossible to avoid local distortion of the crystal lattice. Formation of complexes minimizes the energy of elastic deformation in the crystal.

The emissive transition between the 2$D$ bands of the $ZnSe$ quantum well is identified the $L_{4}$ band with the lowest energy. The components  L$_{4(1)}$ and L$_{4(2)}$ correspond to light and heavy excitons transitions. The spectral position of the L$_4$ band depends on $B$, because the height of the potential barrier from the side of the $Zn_{0.9}Be_{0.05}Mn_{0.05}Se$ semimagnetic semiconductor changes in the magnetic field. The potential barrier height for the electrons having a spin oriented along $B$ decreases and for the electrons having spin oriented opposite to $B$ increases, if the magnetic field induction increases as a result of the Zeeman splitting of the $C$ and $V$ bands of the $Zn_{0.9}Be_{0.05}Mn_{0.05}Se$ layer.

The conduction bands offset between $Zn_{0.9}Be_{0.05}Mn_{0.05}Se$ and $ZnSe$ exceeds the respective offset between the valence bands of these layers noticeably. In the first approximation, it is therefore reasonable to assume that mainly the position of the valence band edge of the $ZnSe$ quantum well changes as a result of the change of the potential barrier height caused by Zeeman splitting. This approximation is shown by magenta lines in Fig.~\ref{f5}.

Let us concentrate on the field dependence of the $\sigma^{-}$ polarization bands.  In our opinion, it is the most characteristic property of the PL spectra obtained, which provides data about the peculiarities of the interaction between $Mn$ ions and free carriers in the investigated structure. The $L_{3'}$ band shifts to the short-wave range until the magnetic field arrives to the value of about 1.5 T and changes the shift direction for the fields above 1.75 T (Fig.~\ref{f4}~(a)). By comparing these values with the $a1$ structure energy diagram (Fig.~\ref{f5}) it is evident to see that the change of the $L_{3'}$  band behavior occurs in the magnetic field range where the energy level of $Mn$ complex $E^{Mn^{+}}_{ZnBeMnSe}$ in $Zn_{0.9}Be_{0.05}Mn_{0.05}Se$  comes into resonance with the $V$ band edge $E^{V^{+}}_{ZnBeMnSe}$ of the barrier.

There is another characteristic feature of the $L_{3'}$ band behavior in the range of its inverse shift. In the fields above 2 T energy position of the $\sigma^{-}$ polarized $L_{3'}$ band coincides with an energy position of the $\sigma^{+}$ polarized $L_{1}$ band (Fig.~\ref{f4} (a)). At the same time the $L_{3'}$ band intensity in this field range is about 8 to 15 times smaller than that of the $L_{1}$ band. The two peculiarities taken together are the key for understanding the behavior of the $L_{3'}$ luminescence band.

It is known \cite{8} that there are processes of light scattering in a magnetic field with the electrons of the conduction band, which are accompanied by spin-flip phenomena in semiconductors with the degenerate valence band. If a complex energy level comes into resonance with allowed states of the $V$ band of the quantum barrier, these states become degenerate. Hence, the spin-flip processes become considerable in these magnetic fields. As a result, the $\sigma^{-}$ polarized  $E^{C^{-}}_{ZnSe}$ $\rightarrow$ $E^{Mn^{-}}_{ZnBeMnSe}$  transitions attenuate while the spin-flip induced $\sigma^{-}$ polarized emission transitions with energy equal to a gap between $E^{C^{+}}_{ZnBeMnSe}$  and $E^{V^{+}}_{ZnBeMnSe}$ energy levels is activated. Both the energy gaps $E^{C^{-}}_{ZnSe}$ $\rightarrow$ $E^{Mn^{-}}_{ZnBeMnSe}$ and  $E^{C^{+}}_{ZnBeMnSe}$  $\rightarrow$ $E^{V^{+}}_{ZnBeMnSe}$ are overlapping at the magnetic field of 1.75~T(see energy diagram in Fig.~\ref{f5}). Therefore, the zone of increasing $E(L_{3'})$ field dependence below 1.75 T passes smoothly to the decreasing dependence above.

The $L_{4'}$ band displays an additional distinct behavior. An energy level of the $Mn$ complex in $Zn_{0.9}Be_{0.05}Mn_{0.05}Se$ barrier is in resonance with the allowed states of $V$ band of the $ZnSe$ quantum well at any magnetic field (Fig.~\ref{f5}). Hence, conditions are always fulfilled for spin-flip processes provoked by the quantum states of the $Mn$ complex and the $V$ band of the $ZnSe$ well. However, if the $Mn$ complex energy level $E^{Mn^{+}}_{ZnBeMnSe}$  is farther from the valence band edge $E^{V^{+}}_{ZnSe}$  than the $E^{V^{-}}_{ZnSe}$  energy level the $\sigma^{-}$ polarized emission transitions between $E^{C^{-}}_{ZnSe}$  and $E^{V^{-}}_{ZnSe}$  levels dominates. This becomes obvious from the energy diagram Fig.~\ref{f5}. The spin-flip processes become active only after the  $E^{V^{-}}_{ZnSe}$ and $E^{Mn^{-}}_{ZnBeMnSe}$ levels change places in the energy scale.

In conclusion, luminescence studies of the quantum structures based on the $ZnSe$ quantum well with asymmetrical $ZnBeMnSe$ and $ZnBeSe$ barriers have revealed that: (i) the introduction of $Be$ into semimagnetic $ZnMnSe$ matrix causes decrease of the exchange integrals for conductive and valence bands; (ii) formation of the $ZnBeMnSe$/$ZnSe$/$ZnBeSe$ structure leads to appearance of an emission band in the PL spectra which can be extinguished by an extra $ZnBeSe$ barrier introduced between the $ZnBeMnSe$ and $ZnSe$ layers; (iii) there are some peculiarities in magnetic field dependence of PL spectra of $\sigma^{-}$ polarization which may be attributed to spin-flip processes; (iv) energy level of the $Mn$ complex in the $ZnBeMnSe$ layer may be identified as a possible source of spin-flip processes at high magnetic fields.

\end{document}